\documentclass{amsart}
\usepackage{amsfonts}
\usepackage[polish]{babel}

\theoremstyle{plain}
\newtheorem{theorem}{Theorem}[section]

\theoremstyle{remark}
\newtheorem{remark}[theorem]{Remark}
\numberwithin{equation}{section}

\input{tcilatex}

\begin{document}

\thispagestyle{empty}

\begin{center}
{\footnotesize Available at: \texttt{http://publications.ictp.it}}\hfill
IC/2008/051\\[0pt]
\vspace{1cm} United Nations Educational, Scientific and Cultural Organization%
\\[0pt]
and\\[0pt]
International Atomic Energy Agency\\[0pt]
\medskip THE ABDUS SALAM INTERNATIONAL CENTRE FOR THEORETICAL PHYSICS\\[0pt]
\vspace{1.8cm} \textbf{THE FIELD STRUCTURE OF VACUUM, MAXWELL EQUATIONS AND
RELATIVITY THEORY ASPECTS. Part 1}\\[0pt]
..\bigskip \bigskip \bigskip \vspace{1.5cm}

{\small \textit{The authors dedicate this article to one of the mathematical
and physical giants of the XX-th century - \\[0pt]
academician Prof. Nikolai N. Bogolubov in memory of his 100th Birthday with
great appreciation to his brilliant talent and impressive impact to modern
nonlinear mathematics and quantum physics}}\\[0pt]
Anatoliy K. Prykarpatsky\footnote{%
pryk.anat@ua.fm, prykanat@cybergal.com}\\[0pt]
\textit{The Staanislaw Staszic AGH University of Science and Technology, Krak%
\'{o}w 30-059, Poland,\\[0pt]
The Department of Management at the Ivan Franko State Pedagogical
University, \ Drogobych, Lviv region, \ Ukraina \\[0pt]
}

\textit{\bigskip \bigskip \bigskip Nikolai N. Bogolubov Jr.\footnote{%
nikolai\_bogolubov@hotmail.com}\\[0pt]
V.A. Steklov Mathematical Institute of RAN, Moscow, Russian Federation \\[0pt%
]
and \\[0pt]
The Abdus Salam International Centre for Theoretical Physics, Trieste, Italy%
\\[1em]
}
\end{center}

\baselineskip=18pt \bigskip\bigskip\bigskip

\vfill

\begin{center}
MIRAMARE -- TRIESTE\\[0pt]
February 2008\\[0pt]
\end{center}

\vfill
\newpage \setcounter{page}{1} \centerline{\bf Abstract} \medskip

The vacuum structure and its modeling by means of field theoretic tools is
analyzed. The Maxwell equations from the first principles are derived, the
Lorentz force transformations with respect to non-inertial reference frames
is discussed, some new interpretations of the relativity theory are
presented. \newpage

\section{Introduction}

The nature of space-time and surrounding matter objects was and persists to
be \ a one of the most intriguing and challenging problems facing the
mankind and natural scientists \ especially. As we know one of the most
brilliant inventions in physics of XIX-th century was combining of
electricity and magnetism within the Faraday-Maxwell electromagnetism
theory. This theory explained the main physical laws of light propagation in
space-time and posed new questions concerning the nature of vacuum.
Nonetheless, almost \ all attempts aiming to unveil the real state of art of
the vacuum problem appeared to be unsuccessful in spite of new ideas
suggested by Mach, Lorentz, Poincar\'{e}, Einstein and some others
physicists. \ Moreover, the non-usual way of treating the space-time devised
by Einstein, in reality, favored to eclipsing both its nature and the
related physical vacuum origin problems \cite{Fe,Ga,Ma,TW,Ba,BP}, reducing
them to some physically unmotivated formal mathematical principles and
recipes, combined in the well known special relativity theory (SRT). \ The
SRT appeared to be adapted to the only inertial reference systems and faced
with hard problems of the electromagnetic Lorentz forces explanation and
relationships between inertial and gravity forces. The latter was
artificially "dissolved" by means of the well known "equivalence" principle
owing to which the "inertial" mass of a material object was postulated to
coincide with its "gravity" mass.

Simultaneously, the vacuum origin as a problem almost completely disappeared
from the Einstein theory being replaced by the geometrization of the
space-time nature and all related physical phenomena. Meanwhile, the
impressive success of XX-th century quantum physics, especially of quantum
electrodynamics, have demonstrated clearly enough \cite{BS,AGD,Ba} that the
vacuum polarization and electron-positron annihilation phenomena make it
possible to pose new questions about the space-time and vacuum structures,
and further to revisit \cite{Ma,Fa,BD,Lo} the existing points of view on
them.

Below we try to unveil some nontrivial aspects of the real space-time and
vacuum origin problems to derive from the\ natural field theory principles
all of the well known Maxwell electromagnetism and relativity theories
results, to show their relative or only visible coincidence with real
physical phenomena and to feature new perspectives facing the modern
fundamental physics.

\section{The Maxwell electromagnetism theory: new look and interpretation}

We start from the following field theoretical model of the vacuum considered
as some physical reality imbedded into the standard three-dimensional
Euclidean space reference system marked with three spatial coordinates $r\in 
\mathbb{R}^{3},$ endowed with the standard scalar product $<\cdot ,\cdot >,$
and parameterized by means of the scalar temporal parameter $t\in \mathbb{R}%
. $ \ The physical vacuum we will endow with a smooth enough four-vector
potential function $(W,A):\mathbb{R}^{3}\times \mathbb{R}\rightarrow \mathbb{%
R},$ and material objects, imbedded into the vacuum, we will model
(classically here) by means of the scalar density function $\ \rho :\mathbb{R%
}^{3}\times \mathbb{R}\rightarrow \mathbb{R}$ and the vector current density 
$\ J:\mathbb{R}^{3}\times \mathbb{R}\rightarrow \mathbb{R}^{3},$ being also
smooth enough functions.

\begin{enumerate}
\item The \textit{first} field theory principle about the vacuum we accept
sounds as follows: the four-vector function $(W,A):\mathbb{R}^{3}\times 
\mathbb{R}\rightarrow \mathbb{R}$ satisfies the standard continuity
relationship%
\begin{equation}
\frac{1}{c}\frac{\partial W}{\partial t}+<\nabla ,A>=0,  \label{M1.1}
\end{equation}%
where, by definition, $\nabla :=\partial /\partial r$ is the usual gradient
operator.

\item The \textit{second} field theory principle we accept is a dynamical
relationship on the scalar potential component $W:\mathbb{R}^{3}\times 
\mathbb{R}\rightarrow \mathbb{R}:$%
\begin{equation}
\frac{1}{c^{2}}\frac{\partial ^{2}W}{\partial t^{2}}-\nabla ^{2}W=\rho ,
\label{M1.2}
\end{equation}%
meaning the linear law \ of the small vacuum uniform and isotropic
perturbations propagation in the space-time, understood here, evidently, as
a first (linear) approximation in the case of weak enough fields.

\item The \textit{third }principle is similar to the first one and means
simply the \ natural continuity condition for the density and current
density functions:%
\begin{equation}
\frac{\partial \rho }{\partial t}+<\nabla ,J>=0.  \label{M1.3}
\end{equation}
\end{enumerate}

We need to note here that the vacuum field perturbations velocity parameter $%
c>0,$ used above, coincides with the vacuum light velocity as we are trying
to derive successfully from these first principles the well known Maxwell
electromagnetism field equations, to analyze the related Lorentz forces and
\ special relativity relationships. To do this, we first combine equations (%
\ref{M1.1}) and (\ref{M1.2}):%
\begin{equation*}
\frac{1}{c^{2}}\frac{\partial ^{2}W}{\partial t^{2}}=-<\nabla ,\frac{1}{c}%
\frac{\partial A}{\partial t}>=<\nabla ,\nabla W>+\rho ,
\end{equation*}
whence 
\begin{equation}
<\nabla ,-\frac{1}{c}\frac{\partial A}{\partial t}-\nabla W>=\rho .
\label{M1.4}
\end{equation}%
Having put, by definition,%
\begin{equation}
E:=-\frac{1}{c}\frac{\partial A}{\partial t}-\nabla W,  \label{M1.5}
\end{equation}%
we obtain the first material Maxwell equation%
\begin{equation}
<\nabla ,E>=\rho  \label{M1.6}
\end{equation}%
for the electric field $E:\mathbb{R}^{3}\times \mathbb{R}\rightarrow \mathbb{%
R}^{3}.$ Having now applied the rotor-operation $\ \ \nabla \times $ \ to
expression (\ref{M1.5}) we obtain the first Maxwell field equation%
\begin{equation}
\frac{1}{c}\frac{\partial B}{\partial t}-\nabla \times E=0  \label{M1.7}
\end{equation}%
on the magnetic field vector function $B:\mathbb{R}^{3}\times \mathbb{R}%
\rightarrow \mathbb{R}^{3},$ defined as and \ \ 
\begin{equation}
B:=\nabla \times A.  \label{M1.8}
\end{equation}

To derive the second Maxwell field equation we will make use of (\ref{M1.8}%
), (\ref{M1.1}) and (\ref{M1.5}):%
\begin{eqnarray}
\nabla \times B &=&\nabla \times (\nabla \times A)=\nabla <\nabla ,A>-\nabla
^{2}A=  \notag \\
&=&\nabla (-\frac{1}{c}\frac{\partial W}{\partial t})-\nabla ^{2}A=\frac{1}{c%
}\frac{\partial }{\partial t}(-\nabla W-\frac{1}{c}\frac{\partial A}{%
\partial t}+\frac{1}{c}\frac{\partial A}{\partial t})-\nabla ^{2}A=  \notag
\\
&=&\frac{1}{c}\frac{\partial E}{\partial t}+(\frac{1}{c^{2}}\frac{\partial
^{2}A}{\partial t^{2}}-\nabla ^{2}A).  \label{M1.9}
\end{eqnarray}%
We have from (\ref{M1.5}), (\ref{M1.6}) and (\ref{M1.3}) that 
\begin{equation*}
<\nabla ,\frac{1}{c}\frac{\partial E}{\partial t}>=\frac{1}{c}\frac{\partial
\rho }{\partial t}=-\frac{1}{c}<\nabla ,J>,
\end{equation*}%
or 
\begin{equation}
<\nabla ,-\frac{1}{c}\frac{\partial ^{2}A}{\partial t^{2}}-\nabla (\frac{1}{c%
}\frac{\partial W}{\partial t})+\frac{1}{c}J>=0.  \label{M1.10}
\end{equation}%
Making now use of (\ref{M1.1}), from (\ref{M1.10}) we obtain that%
\begin{eqnarray}
&<&\nabla ,-\frac{1}{c^{2}}\frac{\partial ^{2}A}{\partial t^{2}}-\nabla (%
\frac{1}{c}\frac{\partial W}{\partial t})+\nabla ^{2}A+\nabla \times (\nabla
\times A)+\frac{1}{c}J>=  \notag \\
&=&<\nabla ,-\frac{1}{c^{2}}\frac{\partial ^{2}A}{\partial t^{2}}+\nabla
^{2}A+\frac{1}{c}J>=0.  \label{M1.11}
\end{eqnarray}%
Thereby, equation (\ref{M1.11}) yields%
\begin{equation}
\frac{1}{c^{2}}\frac{\partial ^{2}A}{\partial t^{2}}-\nabla ^{2}A=\frac{1}{c}%
(J+\nabla \times S)  \label{M1.12}
\end{equation}%
or for some smooth vector function $S:\mathbb{R}^{3}\times \mathbb{R}%
\rightarrow \mathbb{R}^{3}.$ Here we need to note that continuity equation (%
\ref{M1.3}) is defined, concerning the current density vector $J:\mathbb{R}%
^{3}\times \mathbb{R}\rightarrow \mathbb{R}^{3},$ up to a vorticity
expression, that is $J\simeq J+\nabla \times S$ and equation (\ref{M1.12})
finally can be written down as 
\begin{equation}
\frac{1}{c^{2}}\frac{\partial ^{2}A}{\partial t^{2}}-\nabla ^{2}A=\frac{1}{c}%
J.  \label{M1.13}
\end{equation}%
Having substituted now (\ref{M1.13}) into (\ref{M1.9}) we obtain the second
Maxwell field equation

\begin{equation}
\nabla \times B-\frac{1}{c}\frac{\partial E}{\partial t}=\frac{1}{c}J.
\label{M1.14}
\end{equation}%
In addition, from (\ref{M1.8}) one also finds the no-magnetic charge
relationship%
\begin{equation}
<\nabla ,B>=0.  \label{M1.15}
\end{equation}

Thus, we have derived all the Maxwell electromagnetic field equations from
our three main principles (\ref{M1.1}), (\ref{M1.2}) and (\ref{M1.3}). The
success of our undertaking will be more impressive if we adapt our results
to those following from the well known relativity theory in the case of
point charges or masses. Below we will try to demonstrate the corresponding
derivations based on some completely new physical conceptions of the vacuum
medium first discussed in \cite{Re}.

\begin{remark}
It is interesting to analyze a partial case of the first field theory vacuum
principle (\ref{M1.1}) when the following conservation law for the scalar
potential field function $W:\mathbb{R}^{3}\times \mathbb{R}\rightarrow 
\mathbb{R}$ holds:%
\begin{equation}
\frac{d}{dt}\int_{\Omega _{t}}Wd^{3}r=0,  \label{M1.16}
\end{equation}%
where $\Omega _{t}\subset \mathbb{R}^{3}$ is ny open domain in space $%
\mathbb{R}^{3}$ with the smooth boundary $\partial \Omega _{t}$ for all $%
t\in \mathbb{R}$ and $d^{3}r$ is the standard volume measure in $\mathbb{R}%
^{3}$in a vicinity of the point $r\in \Omega _{t}.$
\end{remark}

Having calculated expression (\ref{M1.16}) we obtain the following
equivalent continuity equation 
\begin{equation}
\frac{1}{c}\frac{\partial W}{\partial t}+<\nabla ,\frac{v}{c}W>=0,
\label{M1.17}
\end{equation}%
where $\nabla :=\nabla _{r}$ and $v:=dr/dt$ is the velocity vector of a
vacuum medium perturbation at point $r\in \mathbb{R}^{3}$ carrying the field
potential quantity $W.$ Comparing now equations (\ref{M1.1}), (\ref{M1.17})
and using equation (\ref{M1.3}) we can make the suitable identifications: 
\begin{equation}
A=\frac{v}{c}W,\text{ \ \ \ \ }J=\rho v,  \label{M1.18}
\end{equation}%
well known from the classical superconductivity theory \cite{Ga}. Thus, we
face with a new physical interpretation of the conservative electromagnetic
field theory when the vector \ potential $A:\mathbb{R}^{3}\times \mathbb{R}%
\rightarrow \mathbb{R}^{3}$ completely determines via expression (\ref{M1.18}%
) by the scalar field potential function $W:\mathbb{R}^{3}\times \mathbb{R}%
\rightarrow \mathbb{R}.$ It is also evident that all of the Maxwell
electromagnetism filed equations derived above hold too in the case (\ref%
{M1.18}), as it was first demonstrated in \cite{Re} (but with some
mathematical inaccuracies).

Consider now the conservation equation (\ref{M1.16}) jointly with the
related integral momentum conservation relationship 
\begin{equation}
\frac{d}{dt}\int_{\Omega _{t}}(\frac{vW}{c^{2}})d^{3}r=0,~~~\Omega
_{t}|_{t=0}=\Omega _{0},  \label{M1.19}
\end{equation}%
where as above $\Omega _{t}\subset \mathbb{R}^{3}$ is for any time $t\in 
\mathbb{R}$ an open domain with the smooth boundary $\partial {\Omega _{t}}, 
$ whose evolution is governed by the equation 
\begin{equation}
dr/dt=v(r,t)  \label{M1.20}
\end{equation}%
for all $x\in \Omega _{t}$ and $t\in \mathbb{R},$ as well as by the initial
state of the boundary $\partial {\Omega }_{0}.$ As a result of relationship (%
\ref{M1.19}) one obtains the new continuity equation 
\begin{equation}
\frac{d(vW)}{dt}+vW<\nabla ,{v}>=0.  \label{c11}
\end{equation}%
Making now use of (\ref{M1.17}) in the equivalent form 
\begin{equation*}
\frac{dW}{dt}+W<\nabla ,{v}>=0,
\end{equation*}%
we obtain finally a very interesting local conservation relationship \ 
\begin{equation}
dv/dt=0 \   \label{c13}
\end{equation}%
on the vacuum matter velocity $v=dr/dt,$ holding for all values of the time
parameter $t\in \mathbb{R}.$ As it is easy to observe, the obtained
relationship completely coincides with the well known hydrodynamic equation 
\cite{MC} of ideal compressiable liquid without any external exertion, that
is any external forces and field "pressure" \ are equal identically to zero.
We received a natural enough result that the propagation velocity of the
vacuum field matter is constant and equals exactly $v=c,$ that is the light
velocity in the vacuum, if to recall the starting wave equation (\ref{M1.2})
owing to which the small vacuum field matter perturbations propagate in the
space with the light velocity.

\section{Special relativity and dynamical field equations}

From classical electrodynamics we know that the main dynamical relationship
relates the particle mass acceleration with the Lorentz force which strongly
depends on the absolute \bigskip charge velocity. For the electrodynamics to
be independent on the reference system physicists were forced to reject the
Galilean transformations and replace them with the artificial Lorentz
transformations. This resulted later in the Einstein relativity theory which
has partly reconciled the problems concerned with deriving true dynamical
equations for a charged point particle.

We will now start from the scalar field vacuum medium function $W:\mathbb{R}%
^{3}\times \mathbb{R}\rightarrow \mathbb{R}$ in the conservation condition
case (\ref{M1.16}) discussed above. This means, obviously, that the vacuum
medium field vector potential $A:\mathbb{R}^{3}\times \mathbb{R}\rightarrow 
\mathbb{R}^{3},$ charge and current densities $(\rho ,J):\mathbb{R}%
^{3}\times \mathbb{R}\rightarrow \mathbb{R}^{3}$ $\times \mathbb{R}$ are
related owing to expressions (\ref{M1.18}).

Consider now jointly vacuum field medium conservation equations (\ref{M1.17}%
) and (\ref{M1.2}) at the density $\rho =0:$%
\begin{eqnarray}
-\frac{1}{c^{2}}\frac{\partial ^{2}W}{\partial t^{2}} &=&\frac{1}{c^{2}}%
\frac{\partial }{\partial t}(-\frac{\partial W}{\partial t})=\frac{1}{c^{2}}%
\frac{\partial }{\partial t}(<\nabla ,vW>)=  \notag \\
&=&<\nabla ,\frac{\partial }{\partial t}(\frac{Wv}{c^{2}})>=-<\nabla ,\nabla
W>.  \label{M2.1}
\end{eqnarray}%
From relationship (\ref{M2.1}) one follows that%
\begin{equation}
\frac{\partial }{\partial t}(\frac{Wv}{c^{2}})+\nabla W=\nabla \times F,
\label{M2.2}
\end{equation}%
where $F:\mathbb{R}^{3}\times \mathbb{R}\rightarrow \mathbb{R}^{3}$ is some
smooth function, which we put, by definition, to be zero owing to the 
\textit{a priori} assumed vortexless vacuum medium dynamics. So, our
dynamical equation on the vacuum medium scalar field function $W:\mathbb{R}%
^{3}\times \mathbb{R}\rightarrow \mathbb{R}$ looks like 
\begin{equation}
\frac{\partial }{\partial t}(\frac{Wv}{c^{2}})+\nabla W=0.  \label{M2.3}
\end{equation}

Consider now a charged point particle $q$ in a space point $%
r=R(t):=R_{0}+\int_{0}^{t}udt\in \mathbb{R}^{3},$ depending on time
parameter $t\in \mathbb{R}$ and position $R_{0}\in $ $\mathbb{R}^{3}$ at
intial point $t=0\mathbf{.}$ Since the vacuum medium field is described by
means of the potential field function $W:\mathbb{R}^{3}\times \mathbb{R}%
\rightarrow \mathbb{R},$ which is naturally disturbed by the charged
particle $q,$ we will model this fact approximately as the following
resulting functional relationship:%
\begin{equation}
W(r,t)=\tilde{W}(r,R(t))  \label{M2.4}
\end{equation}%
for some scalar function $\tilde{W}:\mathbb{R}^{3}\times \mathbb{R}%
^{3}\rightarrow \mathbb{R}.$ This function must satisfy equation (\ref{M1.17}%
), that is 
\begin{equation}
<\frac{\partial \tilde{W}}{\partial R},u>+<\nabla ,\tilde{W}v>=0.
\label{M2.5}
\end{equation}%
As we are interested in the function $\tilde{W}:\mathbb{R}^{3}\times \mathbb{%
R}\rightarrow \mathbb{R}$ as $r\rightarrow R(t)\in \mathbb{R}^{3},$ where
the charged point particle is located, we obtain from (\ref{M2.5}) that 
\begin{equation*}
<\frac{\partial \tilde{W}}{\partial R}+\frac{\partial \tilde{W}}{\partial r}%
,u>|_{r\rightarrow R(t)}+\tilde{W}<\nabla ,v>|_{r\rightarrow R(t)}=0,
\end{equation*}%
giving rise to the relationship 
\begin{equation}
\frac{\partial \tilde{W}}{\partial R}=-\frac{\partial \tilde{W}}{\partial r}
\label{M2.6}
\end{equation}%
as $r\rightarrow R(t),$ since $v|_{r\rightarrow R(t)}\rightarrow
dR(t)/dt:=u(t)$ and $<\nabla ,v>|_{r\rightarrow R(t)}\rightarrow <\nabla
,u(t)>=0$ for all $t\in \mathbb{R}.$

Returning now to equation (\ref{M2.3}) we can write down, owing to (\ref%
{M2.6}), that 
\begin{equation}
\begin{array}{c}
\frac{1}{c^{2}}\left. \left( \frac{\partial \tilde{W}}{\partial t}v+\tilde{W}%
\frac{\partial v}{\partial t}\right) \right\vert _{r\text{ }\rightarrow
R(t)}=\frac{1}{c^{2}}\left. \left( -<\frac{\partial \tilde{W}}{\partial r}%
,v>v+\tilde{W}\frac{\partial v}{\partial t}\right) \right\vert
_{r\rightarrow R(t)}= \\ 
=\frac{1}{c^{2}}\left. \left( <\frac{\partial \tilde{W}}{\partial R},u>u+%
\tilde{W}\frac{du}{dt}\right) \right\vert _{r\rightarrow R(t)}=\frac{1}{c^{2}%
}\frac{d}{dt}(\bar{W}u)=-\left. \frac{\partial \tilde{W}}{\partial r}%
\right\vert _{r\rightarrow R(t)}=\frac{\partial \bar{W}}{\partial R},%
\end{array}
\label{M2.7}
\end{equation}%
where we put, by definition, $\bar{W}:=\tilde{W}(r,R(t))|_{r\rightarrow
R_{0}}.$ Thus, we obtained from (\ref{M2.7}) that the function $\bar{W}:%
\mathbb{R}^{3}\rightarrow \mathbb{R}$ satisfies the determining dynamical
equation 
\begin{equation}
\frac{d}{dt}(-\frac{\bar{W}}{c^{2}}u)=-\frac{\partial \bar{W}}{\partial R}
\label{M2.8}
\end{equation}%
at point $R(t)\in \mathbb{R}^{3}$ of the point charge $q$ location.

Now we need to proceed with our calculations ahead and would like to
interpret the quantity $-\frac{\bar{W}}{c^{2}}$ as the real "dynamical" mass
of our point charge $q$ at point $R(t)\in \mathbb{R}^{3},$ that is 
\begin{equation}
m:=-\frac{\bar{W}}{c^{2}}.  \label{M2.9}
\end{equation}%
Then, using (\ref{M2.9}) we can rewrite equation (\ref{M2.8}) as 
\begin{equation}
\frac{dp}{dt}=-\frac{\partial \bar{W}}{\partial R},  \label{M2.10}
\end{equation}%
where the quantity $p:=mu$ has the natural momentum interpretation.

The obtained equation (\ref{M2.10}) is very interesting from the dynamical
point of view. Really, from equation (\ref{M2.10}) we obtain that 
\begin{equation}
<u,\frac{d}{dt}(mu)>=c^{2}<\frac{\partial m}{\partial R},u>=c^{2}\frac{dm}{dt%
}.  \label{M2.11}
\end{equation}%
As a result of (\ref{M2.11}) we easily derive, following \cite{Re}, the
conservative relationship%
\begin{equation}
\frac{d}{dt}\left( m\sqrt{1-\frac{u^{2}}{c^{2}}}\right) =0  \label{M2.12}
\end{equation}
for all $t\in \mathbb{R}.$ Thereby, the quantity 
\begin{equation}
m\sqrt{1-\frac{u^{2}}{c^{2}}}=m_{0}  \label{M2.13}
\end{equation}%
is constant for all $t\in \mathbb{R},$ giving rise to the well known
relativistic expression for the mass of a point particle:%
\begin{equation}
m=\frac{m_{0}}{\sqrt{1-\frac{u^{2}}{c^{2}}}}.  \label{M2.14}
\end{equation}%
As we can see, the point particle mass $m$ depends, in reality, not on the
coordinate $R(t)\in \mathbb{R}^{3}$ of the point particle $q,$ but on its
velocity $u:=dR(t)/dt.$ Since the field potential $\bar{W}:\mathbb{R}%
^{3}\rightarrow \mathbb{R}$ \ consists of two parts 
\begin{equation}
\bar{W}=\bar{W}_{0}+\Delta \bar{w},  \label{M2.15}
\end{equation}%
where $\bar{W}_{0}:\mathbb{R}^{3}\rightarrow \mathbb{R}$ is constant and
responsible for the external influence of all long distant objects in the
Universe upon the point particle $q$ and $\Delta \bar{w}:\mathbb{R}%
^{3}\rightarrow \mathbb{R}$ is responsible for the local field potential
perturbation by the point charge $q$ and its closest ambient \ neighborhood.
Then, obviously, 
\begin{equation}
\Delta m:=m-m_{0}=-\Delta \bar{w}/c^{2}  \label{M2.16}
\end{equation}%
is the strictly dynamical mass component belonging to the point particle $q.$
Moreover, since the full momentum $p=mu$ satisfies equation (\ref{M2.10}),
one can easily obtain that the quantity 
\begin{equation}
\bar{W}^{2}-p^{2}c^{2}=E_{0}^{2}  \label{M2.17}
\end{equation}
is not depending on time $t\in \mathbb{R},$ that is $dE_{0}/dt=0,$ where $%
E_{0}:=m_{0}c^{2}.$ The result (\ref{M2.17}) demonstrates us the important
property of the energy essence: the point particle $q$ is, in reality,
endowed with the only dynamical energy $\Delta E:=\Delta mc^{2}.$ Concerning
the so called "internal" particle energy $E_{0}=m_{0}c^{2}$ we see that it
has nothing to do with the real particle energy, since its origin is
determined completely owing to the long distant objects of the Universe and
could not be used for any physical processes, contrary to the known Einstein
theory statements about a "huge" internal energy stored inside the particle
mass. Equivalently, the Einstein theory statement about the "equivalence" of
the mass and the "internal" energy of particle appears to be senseless,
since the main part of the field potential function $\bar{W}:\mathbb{R}%
^{3}\rightarrow \mathbb{R}$ at the location point of the particle $q$ is
constant and owes to the long distant objects in the Universe, which
obviously can not be used for so called "practical applications".

Nonetheless, we \ have observed above, as a by-product, the well known
"relativistic" effect of the particle mass growth depending on the particle
velocity in the form (\ref{M2.14}). As it was already mentioned in \cite{Re}
this "mass growth" is, in reality, completely of dynamical nature and is not
a consequence of the Lorentz transformations, as it was stated within the
Einstein SRT. Moreover, we can state that all of so called "relativistic"
effects have also nothing to do with both the mentioned above Lorentz
transformations and with such artificial "effects" as length "shortening"
and time "slowing". There is also no reasonable cause to identify the
particle mass with its real energy and vice versa. Concerning the
interesting physical effect called particles "annihilation" we need here to
stress that it has also nothing to do with the transformation of particles
masses into energy. The field theoretical explanation of this phenomenon
consists in creating their very special bonding state, whose interaction
with ambient objects is vanishing. As a result the visible inertial or
dynamical mass of this bound state is also zero, what the experiment shows,
and nothing else. Inversely, if an intensive enough gamma-quant meets such a
bound state of two particles, it can break them back into two separate
particles, what the experiment also shows to happen. Here we can recall a
similar analogy borrowed from the modern quantum physics of infinite
particle systems described by means of the\ second quantization scenario 
\cite{Fo,BP} suggested in 1932 by V. Fock. Within this scenario there also
realize creating-annihilation effects which are present owing to the
inter-particle interaction forces. Moreover, as we know from the modern
superfluidity and superconductivity theories within this picture one can
describe special bound states of particles, so called "Couper pairs", whose
interaction to each other completely vanishes and whose combined mass
strongly\ differs from the sum of the suitable components and equals the so
called "effective" compound mass, depending strongly on the potential field
intensity inside the superfluid or superconductor matter.

\section{Relativity principles revisiting}

It is a well known fact that the Einstein special relativity theory is
applicable only for physical processes related to each other by means of the
inertial reference systems, moving with constant velocities. In this case
one can make use of the Lorentz transformations and calculate the components
of suitable four-vectors and the resulting mass growth of particles owing to
formula (\ref{M2.11}). A nontrivial problem arises when we wish to analyze
these quantities with respect to non-inertial reference systems moving with
some nonzero acceleration. Below we will revisit this problem from the
devised above vacuum field theory scenario and show that the whole "special"
relativity theory emerges as its partial case or by-product and is free of
the artificial "inertial reference systems" problems mentioned above.

Really, our\ vacuum field theory \ structure is described by dynamical
equation (\ref{M2.3}), which we would like to investigate in a neighborhood
of two interacting to each other point particles $q_{1}$ at point $%
R_{1}(t)\in \mathbb{R}^{3}$ and $\ q_{2}$ at point $R_{2}(t)\in \mathbb{R}%
^{3},$ respectively. As it was already done in Section 2 we assume that the\
vacuum potential field function $W:\mathbb{R}^{3}\times \mathbb{R}%
\rightarrow \mathbb{R}$ can be representable as $W=\tilde{W}%
(r;R_{1}(t),R_{2}(t))$ for some function $\tilde{W}:\mathbb{R}^{3}\times 
\mathbb{R}^{3}\times \mathbb{R}^{3}\rightarrow \mathbb{R}$ and all $t\in 
\mathbb{R}.$ Then, based on continuity equation (\ref{M1.17}) we obtain%
\begin{equation}
<\frac{\partial \tilde{W}}{\partial R_{1}},u_{1}>+<\frac{\partial \tilde{W}}{%
\partial R_{2}},u_{2}>+<\frac{\partial \tilde{W}}{\partial r},v>+\tilde{W}%
<\nabla ,v>=0.  \label{M3.1}
\end{equation}%
We will now be interested in the potential field function $\tilde{W}:\mathbb{%
R}^{3}\times \mathbb{R}^{3}\times \mathbb{R}^{3}\rightarrow \mathbb{R}$ in a
vicinity of the relative vector $R(t):=R_{2}(t)-R_{1}(t)\in \mathbb{R}^{3},$
keeping in mind that the interaction between particles $q_{1}$ and $q_{2%
\text{ }}$depends on this relative interparticle distance $R(t)\in \mathbb{R}%
^{3}.$ From (\ref{M3.1}) as the relative distance $r\rightarrow R(t)\in 
\mathbb{R}^{3}$ we derive easily that%
\begin{equation}
\left. \frac{\partial \tilde{W}}{\partial R_{1}}+\frac{\partial \tilde{W}}{%
\partial R_{2}}\right\vert _{r\rightarrow R(t)}=0,\text{ \ \ \ \ }\left. 
\frac{\partial \tilde{W}}{\partial R}+\frac{\partial \tilde{W}}{\partial r}%
\right\vert _{r\rightarrow R(t)}=0.  \label{M3.2}
\end{equation}%
Combining relationships (\ref{M3.2}) with dynamical field equations (\ref%
{M2.3}) we obtain that 
\begin{equation*}
\begin{array}{c}
\frac{1}{c^{2}}\frac{\partial }{\partial t}\left. \left( \tilde{W}v\right)
\right\vert _{r\text{ }\rightarrow R(t)}=\frac{1}{c^{2}}\left. \left( <\frac{%
\partial \tilde{W}}{\partial R_{1}},u_{1}>v+<\frac{\partial \tilde{W}}{%
\partial R_{2}},u_{2}>v+\tilde{W}\frac{\partial v}{\partial t}\right)
\right\vert _{r\rightarrow R(t)}= \\ 
=\frac{1}{c^{2}}\left. \left( <\frac{\partial \tilde{W}}{\partial R}%
,u_{2}-u_{1}>(u_{2}-u_{1})\right) \right\vert _{r\rightarrow R(t)}=-\left. 
\frac{\partial \tilde{W}}{\partial r}\right\vert _{r\rightarrow R(t)}=\left. 
\frac{\partial \tilde{W}}{\partial R}\right\vert _{r\rightarrow R(t)},%
\end{array}%
\end{equation*}%
whence one derives the new dynamical equation 
\begin{equation}
\frac{d}{dt}(-\frac{\bar{W}}{c^{2}}(u_{2}-u_{1}))=-\frac{\partial \bar{W}}{%
\partial R}  \label{M3.3}
\end{equation}%
on the resulting function $\bar{W}:=\tilde{W}|_{r\rightarrow R(t)}.$

Equation (\ref{M3.3}) possesses a very important feature of depending on the
only relative quantities not depending on the reference system. Moreover, we
have not on the whole met the necessity to use other transformations of
coordinates different from the Galilean transformations. We mention here
that dynamical equation (\ref{M3.3}) was also derived in \cite{Re} making
use of some not completely true relationships and mathematical
manipulations. But the main corollary of \cite{Re} and our derivation,
saying that equation (\ref{M3.3}) fits for all reference systems, both
inertial and accelerated, appears to be fundamental and give rise to new
unexpected results in the modern electrodynamics. Below we will proceed to
one of very important relativity physics aspect, concerned with the well
known Lorentz force expression measuring the action exerted by external
electromagnetic field on a charged point particle $q$ at space point $%
R_{2}(t)\in \mathbb{R}^{3}$ for any time moment $t\in \mathbb{R}.$

To do this we put, owing to the vacuum field theory, that the resulting
potential field function $\bar{W}:\mathbb{R}^{3}\rightarrow \mathbb{R}$ can
be representable in a vicinity of the charged point particle $q$ as 
\begin{equation}
\bar{W}=\bar{W}_{0}+q\varphi ,  \label{M3.3a}
\end{equation}%
where $\varphi :\mathbb{R}^{3}\rightarrow \mathbb{R}$ is a suitable local
electromagnetic field potential and $\bar{W}_{0}:\mathbb{R}^{3}\rightarrow 
\mathbb{R}$ is a constant vacuum field potential owing to the particle
interaction with the external distant Universe objects. Then, having
substituted (\ref{M3.3a}) into main dynamical field equation (\ref{M3.3}) we
obtain that 
\begin{eqnarray}
\frac{d}{dt}(-\frac{\bar{W}}{c^{2}}u) &=&\frac{d}{dt}(-\frac{\bar{W}}{c^{2}}%
v)-\nabla \bar{W}=-\nabla \bar{W}+\frac{\partial }{\partial t}(-\frac{\bar{W}%
}{c^{2}}v)+<u,\nabla >(-\frac{\bar{W}}{c^{2}}v)=  \notag \\
&=&-\nabla \bar{W}+\frac{1}{c}\frac{\partial }{\partial t}(-\frac{\bar{W}}{c}%
v)-u\times (v\times \nabla \frac{\bar{W}}{c^{2}})-<u,v>\nabla \bar{W}= 
\notag \\
&=&-\nabla \bar{W}(1+\frac{<u,v>}{c^{2}})+\frac{1}{c}\frac{\partial }{%
\partial t}(-\frac{\bar{W}}{c}v)-\frac{1}{c^{2}}u\times (v\times \nabla \bar{%
W})=  \notag \\
&=&-q\nabla \varphi (1+\frac{<u,v>}{c^{2}})-\frac{q}{c}\frac{\partial }{%
\partial t}(\frac{\varphi }{c}v)+\frac{q}{c}u\times (\nabla \times \frac{%
\varphi v}{c})=  \notag \\
&=&-q\nabla \varphi (1+\frac{<u,v>}{c^{2}})-\frac{q}{c}\frac{\partial A}{%
\partial t}+\frac{q}{c}u\times (\nabla \times A),  \label{M3.3b}
\end{eqnarray}%
where we denoted $u:=u_{2},$ \ $v:=u_{1},$ $\nabla :=\partial /\partial
R_{2}=\partial /\partial R$ and $A:=\varphi v/c,$ being the related magnetic
potential. Since we have already shown that the Lorentz force 
\begin{equation*}
F:=\frac{d}{dt}(-\frac{\bar{W}}{c^{2}}u)=\frac{d}{dt}\left( \frac{m_{0}}{%
\sqrt{1-\frac{u^{2}}{c^{2}}}}\right)
\end{equation*}%
is given by expression (\ref{M3.3b}), it can be rewritten down in the form 
\begin{eqnarray}
F &=&\frac{d}{dt}\left( \frac{m_{0}}{\sqrt{1-\frac{u^{2}}{c^{2}}}}\right)
=qE+\frac{q}{c}u\times B-\frac{q}{c^{2}}\nabla \varphi <u,v>=  \label{M3.4}
\\
&=&qE+\frac{q}{c}u\times B-\frac{q}{c}\nabla <u,A>,  \notag
\end{eqnarray}%
which was derived also in \cite{Re} and where we put, by definition, $%
E:=-\nabla \varphi -\frac{1}{c}\frac{\partial A}{\partial t},$ \ $B:=\nabla
\times A,$ being respectively the suitable electric and magnetic vector
fields.

\section{Conclusion}

The resulting expression (\ref{M3.4}) is almost completely equivalent to the
well known classical Lorentz force $F$ up to the additional "inertial" term 
\begin{equation}
F_{c}:=-\frac{q}{c}\nabla <u,A>,  \label{M3.5}
\end{equation}%
which is absent in the relativistic theory. Namely, owing to the absence of
term (\ref{M3.5}) the classical relativistic Lorentz force expression was
not invariant with respect to any reference frame transformations, except
inertial ones. And, as it was noticed in \cite{Re}, owing only to this fact
the relativistic physics faced with many difficulties during the past
century and the physicists were forced to use the artificial Lorentz
transformations and related with them visible length shortening and time
slowing effects. Moreover, they gave rise to such strange enough and
non-adequate notions as non-Euclidean time-spaces \cite{Ho,Me,Ba1}, black
holes \cite{Da,Gr,Ho,Ba} and some other nonphysical objects. Concerning the
results described above we could state that the vacuum field theory approach
of \cite{Re} to fundamental physical phenomena is really a powerful tool in
hands of researchers, who wish to penetrate into the hidden properties of
the surrounding us Universe. As the microscopical quantum level of
describing the vacuum field matter structure is, with no doubt, very
important, we see the next challenging steps in understanding the
backgrounds of quantum processes from the approach devised in \cite{Re} and
in this work and in deriving new physical relationships, which will help us
to explain the Nature more deeply and adequately.

\section*{Acknowledgments}

Authors are cordially thankful to the Abdus Salam International Centre for
Theoretical Physics in Trieste, Italy, for the hospitality during their
research 2007-2008 scholarships. A.P. is especially appreciated to Profs.
P.I. Holod (Kyiv, UKMA), J.M. Stakhira (Lviv, NUL), B.M. Barbashov (Dubna,
JINR), Z. K"akol (Krak{\'{o}}w, AGH), J. S{\l }awianowski (Warsaw, IPPT) and
Z. Peradzy{\'{n}}ski (Warsaw, UW) for fruitful discussions, useful comments
and remarks. The last but not least thanks belongs to Prof. O. N. Repchenko
for the discussion of some controversial vacuum field theory aspects and
Mrs. Dilys Drilli (Publishing office, ICTP) for professional help in
preparing a manuscript for publication.


\begin{thebibliography}{99}
\bibitem{Fe} Feynman R. P. Lectures on gravitation. Notes of California
Inst. of Technology, 1971.

\bibitem{BS} Bogolubov N. and Shirkov D. Introduction to the theory of
quantized fields. Interscience, New York, 1959

\bibitem{AGD} Abrikosov A.A., Gorkov L.P. and Dzyaloshinski I.E. Methods of
qunatum field theory in statistical physics. Dover Publication, Inc. New
York, 1975

\bibitem{Ga} de Gennes P.-G. Superconductivity of metals and alloys,
Benjamin, USA, 1964

\bibitem{Ma} Markov M.A. The Mach principle and physical vacuum in general
relativity. Problems of Theoretical Physics. Essays \ dedicated to Nikolai
N. Bogolubov on the occasion of \ his sixtith birthday. \textquotedblleft
Nauka\textquotedblright\ \ Publisher, Moscow, 1969, p. 26-27

\bibitem{Fa} Faddeev L.D. Hamiltonian approach to the gravity. Russian
Physical Surveys, \textquotedblleft Nauka\textquotedblright\ \ Publisher,
Moscow, 1986

\bibitem{Re} Repchenko O. Field physics. Moscow, Galeria Publ., 2005

\bibitem{BD} Brans C.H. and Dicke R.H. Mach's Principle And A Relativistic
Theory Of Gravitation. Phys. Rev., 124, 1961, p. 925

\bibitem{Lo} Logunov A.A. Lectures on relativity theory and gravitation.
\textquotedblleft Nauka\textquotedblright\ \ Publisher, Moscow, 1987

\bibitem{Fo} Fock V.A. Konfigurationraum und zweite Quantelung. Zeischrift
Phys., Bd. 75, 1932, p. 622-647

\bibitem{Ho} t'Hooft G. Introduction to general relativity. Institute for
Theoretical Physics Utrecht University, Princetonplein 5, 3584 CC Utrecht,
the Netherlands, 2002 (www.phys.uu.nl/~thooft/lectures/genrel.pdf)

\bibitem{TW} Taylor E.F. and Wheeler J. A. Spacetime physics. Freeman and
Company, San Francisco and London, 1966

\bibitem{Me} Mermin D.N. It's About Time: Understanding Einstein's
Relativity, Princeton, NJ., Princeton University Press, 2005

\bibitem{Ba1} Barbashov B. M. and Nesterenko V. V. Introduction to the
Relativistic String Theory, World Scientific, Singapore, 1990

\bibitem{Co} Collins H. Gravity's Shadow: The Search for Gravitational
Waves, Chicago: University of Chicago Press, 2004

\bibitem{MC} Marsden J. and Chorin A. Mathematical foundations of the
mechanics of liquid. Springer, New York, 1993

\bibitem{Ba} Barbashov B.M., Efimov G.V. and others. (Editors) Selected
Problems of Modern Physics. Proceedings of the XII-th International
Conference on Selected Problems of Modern Physics. Section 1, Dubna, 2003

\bibitem{BP} Bogolubov N.N. (Jr.), Prykarpatsky A.K. , Golenia J. and Taneri
U. Introductive backgrounds of modern quantum mathematics and application to
nonlinear dynamical systems. Preprint ICTP, Trieste, IC/2007/108
(http://publications.ictp.it)

\bibitem{Da} Damour T. General Relativity and Experiment. Proceedings of the
XI International Congress on Mathematical Physics, Intern. Press, 1995.
Proceedings of the International Conference

\bibitem{Gr} Green B. The elegant Universe. Vintage Books Inc., New York,
1999
\end{thebibliography}
\end{document}